# Scientific Utopia: I. Opening scientific communication

Brian A. Nosek

University of Virginia

Yoav Bar-Anan

Ben-Gurion University

Authors' note: Correspondence concerning this article may be sent to Brian Nosek, nosek@virginia.edu.
The authors have no financial interests concerning the content of this article. Thanks to Mark Brandt, Jamie DeCoster, Yarrow Dunham, Frank Farach, Mike Frank, Carlee Beth Hawkins, Ravi Iyer, Calvin Lai, Matt Motyl, N. Sriram, and Jelte Wicherts for helpful feedback.

Target article for *Psychological Inquiry*

Version 1.2, May 5, 2012

Abstract

Existing norms for scientific communication are rooted in anachronistic practices of bygone eras, making them needlessly inefficient. We outline a path that moves away from the existing model of scientific communication to improve the efficiency in meeting the purpose of public science – knowledge accumulation.  We call for six changes: (1) full embrace of digital communication, (2) open access to all published research, (3) disentangling publication from evaluation, (4) breaking the "one article, one journal" model with a grading system for evaluation and diversified dissemination outlets, (5) publishing peer review, and, (6) allowing open, continuous peer review.  We address conceptual and practical barriers to change, and provide examples showing how the suggested practices are being used already.  The critical barriers to change are not technical or financial; they are social.  While scientists guard the status quo, they also have the power to change it.

Abstract = 143 words

The objective of public science is to build a shared body of knowledge about nature (Goodstein, 2011). To meet this objective, scientists have developed methods and practices that facilitate the acquisition of knowledge. But science is not a particular group or organization. Science is an approach, a set of normative practices, and the process of building and organizing knowledge (http://en.wikipedia.org/wiki/Science). Scientists, the contributors to knowledge accumulation, operate independently, antagonistically, or collaboratively in (mostly) small groups. Thus, the scientific enterprise is a distributed system of agents operating with minimal hierarchical influence.

Open communication among scientists makes it possible to accumulate a shared body of knowledge. No one scientist or scientific body is the arbiter of "truth." Individual scientists or groups make claims and provide evidence for those claims. The claims and evidence are shared publicly so that others can evaluate, challenge, adapt, and reuse the methods or ideas for additional investigation. Truth emerges as a consequence of public scrutiny – some ideas survive, others die. Thus, science makes progress through the open, free exchange of ideas and evidence.[1]

As a key to progress, openness – as embodied by transparency and accessibility – is a central scientific value. Given the distributed nature of scientific practice, a lack of openness reduces the efficiency and veracity of knowledge construction. Ideally, the systems of scientific communication would facilitate openness. When the systems are not operating optimally, the scientific community can redesign them. The core of present day scientific communication is still rooted in the originating 17$^{th}$ century technologies. These technologies do not fully embrace the modern possibilities for openness that would greatly accelerate progress. The question for this article is: How can 21$^{st}$ century scientific communication practices increase openness, transparency, and accessibility in the service of knowledge building?

---

[1] There are certainly many examples of scientific practices that are done in closed and non-sharing circumstances for the development of intellectual property and competitive advantage. These practices are not part of *public* science and are not considered in this article. Also, individual scientists may not share the goals of science as a practice. For example, a scientist's personal goals may be fame or career advancement rather than knowledge building. That is not a problem for science unless the individual's personal goals lead to practices that are in conflict with the goals of science (e.g., faking evidence).

In this article, we describe changes to scientific communication practices. The changes are cumulative as steps away from the existing reality. We titled this article "Scientific Utopia" in recognition that we present an idealized view. The ideas illustrate inefficiencies in the present, and point toward possibilities for improving on those inefficiencies. While an ideal state is not attainable, it can be the basis for of improving current reality. Our purpose is to provide a practical template for improving the current model. We argue that the barriers to these improvements are not technical or financial; they are social. The barriers to change are a combination of inertia, uncertainty about alternative publication models, and the existence of groups invested in the inefficiencies of the present system. Our ultimate goal is to improve research efficiency by bringing scientific communication practices closer to scientific values.

## The Present of Scientific Communication

Scientists are intimately familiar with the modal model of scientific communication – publishing articles in scientific journals. Here, we summarize the key features of that model to set the stage for potential improvements.

**The standard practice**. A team prepares a report of research they conducted. The report summarizes what they did, what they found, and what they think it means. To influence future research and the development of knowledge, the scientists publish the report in a scientific journal. There are approximately 23,750 scientific journals to choose from (Björk, Roos, & Lauri, 2009). Scientific journals have published more than 50 million scholarly articles since the first in 1665, and more than half of these articles have appeared in the last 25 years (Jinha, 2010). The present rate of publication is more than 1.3 million articles per year.

Several features influence the authors' selection of journal. Journals differ on prestige, acceptance rates, topical area, methodological emphasis, report format, readership, length, and publication frequency. Researchers make calculated decisions based on these and other factors to get their research published, promote their careers, and maximize the likelihood that their research will have impact. A common strategy is to submit the article to the most prestigious outlet in the relevant content area that might publish it. However, factors that delay publication are also influential, such as the review lag – how long it takes the journal to decide whether to publish the article, publication lag – how long it takes the journal to print the article in one of its issues after accepting it, and patience lag – how long it takes the authors to become exhausted by the publishing process.

Once submitted, the report is assessed by an editor, an established scientist herself who takes on the role for a limited term. The editor evaluates the report for relevance and publication potential in the journal. If the editor does not reject it immediately, she then solicits reviews from one to five experts in the field to judge its suitability for publication. Those reviewers are selected based on *ad hoc* decision-making by the editor – relevance to content area, known to be reliable or high-quality reviewers, or people she has not asked recently. Potential reviewers accept or decline invitations based on their interest in the article, indebtedness to the editor, and availability. Reviews are typically anonymous, almost always completed without compensation, and can be as short as a few sentences or longer than the report itself. The norm is a few paragraphs summarizing the main issues, and a few follow-up comments regarding minor questions or concerns.

Editors set deadlines for reviews that are sometimes adhered to by the reviewers. Short manuscripts tend to have a faster review process than longer manuscripts. The editor compiles the peer reviews, solicits additional reviews if the present ones are not sufficient, and then renders a decision. The editor has discretionary authority on whether to accept, reject, or invite a revision of the submission. However, it is unusual to make a decision opposing the peer reviewers if they are unanimous, and lack of unanimity is common (Bornmann, Mutz, & Daniel, 2010; Cicchetti, 1990; Marsh & Ball, 1989; Petty, Fleming & Fabrigar, 1999; Peters & Ceci, 1982). The editor then sends a response to the initial author that includes the peer reviews as well as the editor's summary of those reviews. The editor makes a decision to accept, reject, or ask the author to revise and resubmit the manuscript. Invitations to revise the manuscript do not guarantee publication. If the authors submit a revision, the editor may make a decision herself based on the responsiveness to the prior reviews, send it back to one or more of the previous reviewers, or send the manuscript out to new reviewers for additional assessment. This process repeats until the editor accepts or rejects the manuscript or the researchers tire of revising. There is high variability in acceptance rates across journals, some having rates below 10% and others accepting most of the submissions. There are also disciplinary differences (Cole, 1983). For example, rejection rates of 70%-90% are typical in the social sciences, whereas rejection rates of 20%-40% are typical in the physical sciences (e.g., Hargens, 1988; Zuckerman & Merton, 1971).

If the manuscript is rejected, the researchers decide whether to try another journal, and, if so, which one. The same manuscript may be reviewed by multiple teams at multiple journals before acceptance. Once accepted, the manuscript is typeset and copy-edited. The article is then placed in the production queue and scheduled for publication in a future issue. Journals vary in time between issues (from weekly to yearly) and in the length of the issues (from a few articles to more than one hundred). The length and frequency of journal publication is determined by the publisher. The publisher sells journal subscriptions to individuals and institutions. The primary purchasers of subscriptions are university libraries who provide access to the journal for the members of their institution.

**A case study of one laboratory's scientific communication**. While there is substantial data on the acceptance rates and publication timeline for individual journals, there are no known investigations tracking the path of individual manuscripts for publication across journals. A case study of manuscripts authored or co-authored by the first author of the present article may provide some insight into publication practices. Table 1 shows Nosek's unsolicited manuscript submissions to scientific journals that have been in at least one round of peer review.[2] It includes, among other indicators, each manuscript's original submission date and journal, the journal's impact factor (IF), outcome of the editorial process, total number of journals that considered the manuscript, and the total number of days between original submission and the appearance of the article in print.

From July 1999 through April 2012, a total of 62 unsolicited manuscripts were submitted to at least one journal for potential publication. Of those, 39 were published (63%), and five are *in press* (8%). These manuscripts were submitted to an average of 2.0 journals each before being accepted for publication (23 to one journal, 12 to two journals, 12 to three journals, 2 to four journals, 1 to five journals, and 2 to six journals). Of the 39 manuscripts in print, the average time between original submission date and appearance in print was 677 days (median = 564 days). Removing the three articles (8%) that were published in digital-only journals changes that average to 721 days (median = 607).

Of the remaining 18 manuscripts, 14 (23%) were actively in review or revisions for potential publication and 4 (6%) were "in stasis" with no active efforts by the authors to publish the manuscript. These manuscripts were submitted to 2.1 journals each on average (7 to one journal, 5 to two journals, 4 to three journals, 1 to four journals, and 1 to five journals). For these manuscripts, as of 5/3/2012, it had been 1297 days (3.5 years) on average since the original submission date (Median = 568). Excluding the four articles "in stasis" reduces the average time since original submission to 649 days. Overall, 37% (23/62) of the manuscripts were accepted at the first journal to which they were submitted, and 52% (32/62) were submitted to at least two journals.

---

[2] Solicited articles, even if they are peer-reviewed, have a very different publication history. If nothing else, editors are motivated to accept articles that they solicited. In this case example, Nosek had 19 manuscripts solicited by a journal editor for a special issue, as a commentary or review, or for another purpose. 18 (95%) of those were accepted by the original soliciting journal. Also, editors very rarely decline invited chapters for books, articles for encyclopedias, or reports for popular press outlets (case study = 21/21, 100% acceptance rate). While relatively frequent in this case study, *publication-by-direct-invitation* is not addressed in this article.

For articles that have appeared in print, Table 1 also shows each articles citation impact with three indicators: (1) total times the article was cited according to Google Scholar (http://scholar.google.com/), (2) the average number of citations per year since publication, and (3) the article impact factor (IF; source = ISI to match the journal IF database computation).[3] We correlated article impact with indicators of the publishing process in Table 2. Not surprisingly, date of publication correlated strongly with citation impact. Articles in print for a longer period of time accumulated more citations. More interesting is the fact that none of the three indicators of citation impact correlated significantly with the number of journals attempted or the number of days required to publish the article. The citation impact indices did not correlate with the impact factor of the first journal submitted, but did show a positive correlation with the impact factor of the publishing journal. Those correlations estimate that 12%-18% of an article's citation impact can be predicted by the journal in which it appears. It is not clear from these data if this is due to peer review sorting articles into journals by their importance, a self-fulfilling prophecy (i.e., articles in higher prestige/impact outlets are cited more because they are in higher prestige/impact outlets), or both. It is notable that 19 of the 21 available article IFs were greater than the IF's of the first and publishing journals. That suggests most articles exceeded the average article impact for the first journal to which they were submitted.

Whether or not this case study of publication patterns is typical is unknown,[4] but it does reveal wide diversity in the fates of manuscripts submitted for scientific publication. In particular, a majority of articles are reviewed by multiple editorial teams, and the mean time from original manuscript to publication (or to present if not yet published) is more than 2.3 years. We will return to the case study as illustration of important features of the present scientific communication process throughout the article.

---

[3] The article IF is calculated similarly to the journal IF. The article IF is the average number of times the article was cited per year for the two years following its publication (e.g., the average number of citations in 2010 and 2011 for an article published in 2009). The journal IF is the number of times cited in a year for the articles published in the two years preceding (i.e., the average number of citations in 2011 for articles published in 2009 and 2010). To the extent that the journal IF is steady across time, the average article IF should be similar to the journal IF for any particular journal.

[4] Social psychologist, Daryl Bem, anchors one extreme. Over the course of his long career and provocative articles, all of his submitted manuscripts were eventually accepted at the first journal to which he submitted them (Bem, personal communication). We are confident (hopeful?) that Bem is more unusual than Nosek.

## Inefficiencies in Scientific Communication

The focus of the present article is on scientific communication, particularly communication through the scientific journal article. The present system is obviously effective in that science happens and knowledge accumulates. However, there are many ways in which the standard practice does not operate at maximum efficiency in a digital era. The general problems of present scientific communication include:

**A.** *No communication*: Researchers do not always write-up what they did or learned, and some written reports are never published. Rosenthal (1979) named this the "file-drawer effect," which can be very costly for knowledge accumulation. In particular, normative publishing practices make obtaining positive (i.e., significant) results a near necessity for publication (Fanelli, 2010, 2012; Sterling, 1959, 1995). As a result, negative (i.e., nonsignificant) results are much more likely to end up in the file drawer (Greenwald, 1975). This can result in an inflated false-positive rate in the published literature (Ioannidis, 2005), misestimation of meta-analytic effects (Begg, 1994; Egger et al., 1997; Rothstein, Sutton, & Borenstein, 2005; Sutton, Duval, Tweedie, Abrams, & Jones, 2000), and confirmation bias in knowledge building. It is not uncommon for colleagues working in the same field to discover informally that they had all tried an obvious approach to a problem and did not obtain the anticipated effect. Without communication of the initial attempts, there is substantial resource loss on repeated failures, and missed learning opportunities for why an effect does not appear in those anticipated circumstances (Greenwald, Pratkanis, Leippe, & Baumgardner, 1986). Finally, even positive results can fail to be published if the authors do not write a report, or give up on publishing the report because of competing time demands, colleagues leaving the field, or other distractions. Four manuscripts (6%) in the case study had positive results, but are unpublished and in stasis (9.7 years since original submission on average). The process of conducting the case study reminded the author of the existence of those manuscripts. At least three of them would still make a unique contribution to the literature, and the fourth would minimally affirm a now published result.

**B.** *Slow communication*: The case study illustrates that the average published manuscript did not appear in print for nearly two years after it was written. Because science is cumulative, researchers working on similar problems would benefit from learning about each other's results as soon as possible. However, the review process requires time, particularly when the article is reviewed at multiple journals. And, once accepted, publication lags can extend the appearance date by many months.

**C.** *Incomplete communication*: Published articles do not report everything that was done or found. They cannot do so. Reports of a study's methodology reflect the researchers' best understanding of what is crucial in the design. However, this almost always reflects a qualitative assessment of the key factors of the sample, setting, procedures, measures, and context. In other words, the reported methodology describes what the researcher thinks is important, not necessarily what is actually important.[5]

**D.** *Inaccurate communication*: Errors happen, and almost surely in greater numbers than authors and readers realize (Rosenthal, 1978; Rossi, 1987). Scientists do complicated work, and there are few error-detection mechanisms beyond the authors' own diligence. Reviewers can only catch a minority of possible errors because they are constrained by what is reported, and the effort they invest in the review. The true rate of errors is unknown. One study examined errors in reported test statistics, degrees of freedom and p-values in psychology publications (Bakker &

[5] This communication problem is important but not addressed in the present article.

Wicherts, 2011). They found an 18% error rate, and 15% of articles had a statistical conclusion that was incorrect (i.e., reporting a result as significant that was not, or vice versa; see also Garcia-Berthou & Alcaraz, 2004 and Berle & Starcevic, 2007 for examples in other fields).

**E.** *Unmodifiable communication*: Once published, scientific articles are a static entity. The only opportunities for revision are retraction, publication of errata, or comments on the original article, all of which are extremely rare (Budd, Sievert, Schultz, 1998; Redman, Yarandi & Merz, 2008). But, what is not rare is learning that new analytic approaches, reporting styles, or theoretical interpretations are superior to past approaches. Original reports will persist, even if a later report shows there is a clearly better alternative to examining or understanding the evidence. For example, Bishop (2012) identified significant design and analytic problems in a prominent neuroscience article (Temple et al., 2003) based on recent learning about methodology and analysis in this area (e.g., Ioannidis, 2011; Kriegeskorte, Simmons, Bellgowan, & Baker, 2009; Nieuwehuis, Frostmann, & Wagenmakers, 2011). One of the original authors, Poldrack (2012), agreed with the critique noting how some of the criticisms were of standard practice across laboratories for years of fMRI research. Neurocritic (2012) and Bor (2012) then initiated a substantial discussion among dozens of neuroscientists using this as a case example to decide “How much of the neuroimaging literature should we discard?” After all that, the original article remains in print, unmodified. Notably, the critique, response, and subsequent debate all occurred within a matter of days through science blogs operated by scientists online. In the present system, articles with known deficiencies continue to influence future research because there is little incentive or opportunity to reexamine them.

Ironically, communication is often nonexistent, slow, or incomplete *because* the system tries to save time for the scientific community. Authors, reviewers, and editors are gatekeepers. They serve as

quality- and information-control filters. Effective filters are enormously valuable because the volume of scientific information is overwhelming for any individual scientist. However, the present communication practices are not optimizing these filters and, consequently, produce inefficiency. The key impact of these communication problems is wasted time, effort, and resources.

## The Futures of Scientific Communication

In the following sections we propose changes to improve scientific communication. We intend to demonstrate that (a) the present system is improvable, (b) increasing openness will have benefits to the pace and quality of accumulating knowledge, and (c) updates to filtering mechanisms can both accelerate the growth and improve the quality of the scientific literature. Our overall goal is to promote critical review of the systems of scientific communication, and initiate practical steps toward improving them.

We propose six changes to scientific communication. These changes represent a series of stages, and focus on scientific publication of original research. Improving other parts of the scientific process is addressed elsewhere (e.g., Mathieu et al., 2009; Morin et al., 2012; Nosek, 2012; Reichman, Jones, & Schildhauer, 2011; Wicherts, 2011; Wicherts, Borsboom, & Molenaar, 2006; Yarkoni et al., 2010; Schooler, 2011; Stodden, 2011). The stages are ordered in a cumulative fashion, such that later stages partially depend on the changes at earlier stages. However, it is possible to initiate aspects of later stages prior to complete adoption of earlier ones. For each stage, we describe the change, describe what effect it would have on scientific practices, discuss replies to some objections to why the change would be a bad idea, illustrate how the change could be made using existing examples of the practice, and address some practical barriers to performing the change.

### Stage 1: Full embrace of digital communication

The first change is to replace paper with the Internet as the primary mechanism of scientific communication. The existing “standard practice” of the research process remains intact except that communication occurs digitally and, once accepted, articles move to publication very rapidly. In one sense, the transformation to digital has occurred already. Virtually all scientific journals make their articles available digitally. Most scientists use the Internet as their primary means of acquiring and sharing articles. In another sense, key scientific publishing practices are still based on the constraints of publishing on paper.

Publishers accumulate accepted articles, bundle them into issues, print them on paper, and ship them at regular intervals to institutional and individual subscribers. This practice emerged in the 1665 with publication of the first two scientific journals – one of which is still publishing, the *English Philosophical Transactions of the Royal Society*. This practice, initiated almost 400 years ago, is still the guide for how scientific knowledge is communicated today.

Making the Internet the primary vehicle for scientific communication removes printing and shipping and embraces electronic delivery. It makes unnecessary the concept of an "issue" – articles bundled together and sent to subscribers at systematic intervals. Issues are dysfunctional for digital communication because they introduce an irrelevant publication lag between acceptance and availability. This lag varies from months to years, with the typical range probably being between five and ten months, which is between 20% and 40% of the total time between submission and publication in the case study. When articles are made available digitally, publication lag can be eliminated completely. Articles can be published upon completing the editorial review and copyediting process. Notably, the two case study articles with the shortest time to publication both appeared in digital journals that have no publication lag.

It is encouraging that many journal publishers (e.g., Elsevier, Sage) now make "in press" articles available on-line in advance of publication. Even in these cases, however, issues are still eventually printed. As a consequence, another constraint of paper is retained unnecessarily – page limits. Page limits are an underappreciated constraint on scientific communication. The number of articles that can be accepted at a journal is limited by the number of pages that the publisher is willing to print. Printing costs money. Publishers rationally keep constraints on the number of articles published to maintain a profit margin. If the editorial team accepts more than the publisher will print, then the publication lag gets longer. If the editorial team wants to prevent an unwieldy publication lag, then it must accept fewer articles. Whether the journal receives 100, 1000 or 10000 submissions, the number of articles published will be roughly the same. A common editorial decision letter compliments the authors as having done good science, but regrettably notes that the journal gets so many submissions that many good ones are rejected.

Editors easily accept articles that get universal praise and reject articles that get universal disapproval, but most articles are somewhere in between. Inter-rater consistency among article reviewers is low (Bornmann et al., 2010; Marsh & Ball, 1989; Petty et al., 1999; Peters & Ceci, 1982). With just a few reviewers for each submission, this produces a substantial chance element in the publication process (about half of the variability is due to chance, according to one estimate, Whitehurst, 1984). Despite recognition of the chance factors, because of page limits, the editor's default action must be to reject. As a consequence, the same paper can be reviewed by many different journals that are all essentially equivalent on prestige, impact, and quality until the chance factors align sufficiently to earn acceptance. This wastes researchers, reviewers, and editors' time.

In the case study, 20 articles were initially rejected and eventually accepted to journals with known impact factors. Eight of those (40%) were published in a journal that had a similar impact factor (within 1.0) of the originally submitted journal. The article may have improved with revisions across journal submissions (we hope so), but – assuming that the process is rational – such improvement could have been accomplished more efficiently with the original journal, editor and reviewers.

With no page limits, journals can set their publication standards however they wish, and accept as many or as few articles that meet those standards. The size of the journal would be determined by the journal criteria and the quantity and quality of submitted manuscripts. Editors would have more leeway to address the unreliability of the review process and have more flexibility to work with manuscripts at the margin rather than defaulting to rejection. This could reduce (not eliminate) the frequency with which the same “good” article requires multiple rounds of author, editor and reviewer time. Table 2 from the case study illustrated the weak relations among number of journal submissions and time to publication with eventual citation impact.

There are other benefits of digital communication. Printing, shipping, and storage costs are near zero. “Delivery” occurs in multiple ways – a website for continuous access, automated feeds for instant dissemination of new papers relevant to one’s topical interests, and weekly emails to subscribers with highlights of recent articles. Some journals already provide these services. Also, web-based publishing enables improved search and linking capabilities such as adding hyperlinks to citations for immediate article retrieval. Finally, with paper, a researcher must subscribe individually, or have physical access to a subscribing library. With digital communication, access can be just a matter of having an Internet connection. The Research Information Network (Research Information Network, 2008) estimated that converting 90% of publications to electronic-only would save 5% of the costs to publishers and 36% of libraries’ costs to access them (more than $1.6 billion total; see also Houghton, 2009).

**Barriers to change***.* A popular individual concern is that some people like to read articles on paper. Converting to digital distribution does not prevent that. It just requires self-printing, rather than having a publisher print the article along with dozens of others that the individual is not going to read, mailing all articles to all subscribers, and making all readers wait months to read them.

Another possible concern is that journals would need to relax their standards because now they have no page limits. Not at all. Journals can, and should, apply whatever standards they wish. Indeed, some journals use their rejection rates as an indicator of prestige creating value through scarcity and acceptance as an indicator of exclusivity. Standards for acceptance should be based on substantive concerns of the submitted reports; page limits should be irrelevant.

The Internet is a massive equalizer in access to information. Most journals are now online, but still the great majority of the population cannot access it. Access is even highly variable among active scientists depending on which institution employs them. Why is access so limited when the Internet makes it so easy to open it up? That leads us to the next change.

**Stage 2: Open access to all published research**

A closed access publishing model charges subscription fees to readers of the research. An open access model funds publishing with publication fees and then makes the published articles freely available to all potential readers. Presently, the great majority of journals are closed access. But, if someone told you that the publishing model could change to make scientific communication accessible to everyone and simultaneously reduce total publishing costs by $900 million on top of the savings of moving to digital distribution (Research Information Network, 2008), we presume you would think that making the change is a no brainer. We agree. Open access is a financial benefit and a benefit for making information freely available. In Stage 2, we change from a closed access publishing model to an open access one (Harnad, 2003). Everything else in the standard practice remains the same.

Researchers need to have access to the scientific literature in order to be expert at what is known and contribute new knowledge. Practitioners need access to apply the new knowledge. The funding public should have access to know how their money is being spent. Despite this, as of 2012, scientific communication is mostly a closed system. We will briefly explain why, and then describe how this will change (in fact, the transformation is underway).

**Why closed access?** Publishers provide services that scientists and societies could not (or did not want to) do themselves – e.g., typesetting, printing, delivering, subscription management. In exchange for these services, publishers acquire a valuable asset – copyright ownership of the scientific reports.

How do they obtain ownership? The scientists give it to them, for free. Authors are so happy to have their article accepted that they blissfully sign a copyright transfer form sent by the publishers. Then, publishers recoup their investment by closing access to the articles and then selling journal subscriptions to the scientists and their institutions (individual articles can be purchased for $5-$50 depending on the journal). In other words, the funding public, universities, and scientists who produced and pay for the research give ownership of the results to publishers. Then, those with money left over buy the results back from the publishers; the rest are in the dark.

In the paper age, this seemed like a reasonable exchange for the services provided. Individuals had to have physical access to the paper articles. Subscription charges for collating and delivering the paper content made sense. "Open access" was more in control of the libraries storing the paper content than the publishers selling it. In a digital age, where access and delivery are minimal costs, the model seems silly. Institutional journal subscriptions are major costs for university budgets (Houghton, 2009; Houghton et al., 2009). There are more than 23,000 journals, and each is a substantial cost. For example, subscriptions to Elsevier journals alone cost M.I.T. $2 million per year, Purdue $2.3 million per year, and Washington University's School of Medicine $1 million per year.[6] Cutting access to journals is a major cost savings. In 2010, institutions such as Georgia Tech, University of Washington, UCSF, and Oregon State have each dropped hundreds of subscriptions to save hundreds of thousands of dollars per year (Peine, 2011), at the cost of reducing their researchers' access to the literature.

Even with digital publishing there are still costs, but the digital era offers substantial savings opportunities. Besides the benefit of openness for broadening accessibility, the cost-benefit ratio favors open solutions in which authors, funders and institutions pay publishing fees up front over closed solutions in which readers pay subscriptions fees to access (Houghton, 2009; Houghton et al., 2009; Research Information Network, 2008).

---

[6] Sources: http://libraries.mit.edu/sites/scholarly/mit-open-access/open-access-at-mit/mit-open-access-policy/publishers-and-the-mit-faculty-open-access-policy/elsevier-fact-sheet/, http://www.purdueexponent.org/campus/article_19f1031d-0396-5b21-870a-c7f9f6ca91bc.html, and https://becker.wustl.edu/about/news/elsevier-boycott-and-its-relationship-wusm

**How can we open access?** The shift to open access (OA) is in progress.[7] OA journals exist, are gaining awareness and respect in the scientific community, and are sustainable (PLoS, 2010). The non-profit *Public Library of Science* (PLoS; http://plos.org/) is one of the most prominent open access publishers. *PLoS* was founded in 2000 by scientists including Harold Varmus, the Nobel Prize winner and former head of NIH. As of 2012, *PLoS* operated seven OA journals, headlined by two highly selective journals (<10% acceptance rate) *PLoS Medicine* (IF = 13.1) and *PLoS Biology* (IF = 12.2), and *PLoS ONE* (IF = 4.4), a journal publishing articles from any field of science and medicine. At *PLoS ONE* (http://www.plosone.org/), for example, the standard fee for publishing an article was $1350 in 2012. If the researchers do not have grants or university support to cover that fee, the researchers report how much they can pay (as little as $0). The ability to pay has no bearing on the review process or likelihood of acceptance. The editorial board and reviewers are scientist peers, just like other journals. They have no knowledge of whether the researchers are paying and how much. This is critical for avoiding a "pay-to-play" scheme. If accepted, authors pay what they can, and the article is published open access online as soon as the editorial process is complete. There is no print version of *PLoS ONE*.

Funding agencies have recognized that the results of the research they support should be available publicly. For example, despite resistance from some publishers, the National Institutes of Health established PubMed (http://www.ncbi.nlm.nih.gov/pubmed/) as an open access repository of research conducted with NIH funding. Use of repositories for articles published elsewhere is known as "Green OA" whereas OA publishing journals are known as "Gold OA." There are hundreds of repositories, many maintained by universities for their faculty. Also in 2012, one of the largest funders of biomedical research, the Wellcome Trust, adopted an open access policy that requires research they fund to be made available in public repositories, and they provide additional funding to grantees for open access journal publication fees (http://www.wellcome.ac.uk/About-us/Policy/Policy-and-position-statements/WTD002766.htm). Further, in collaboration with the Howard Hughes Medical Institute and the Max Planck Society, the Wellcome Trust launched its own OA journal (http://www.elifesciences.org/) and has committed to underwriting the publishing costs for at least the first few years of operations (http://www.elifesciences.org/about/). Holding the purse strings is a powerful lever to encourage or require open access publishing and to provide the necessary resources to shift away from a subscription-based funding model.

---

[7] Links to in-depth information and resources about Open Access are available here: http://en.wikipedia.org/wiki/Open_access, http://www.earlham.edu/~peters/fos/overview.htm, http://oad.simmons.edu/oadwiki/Main_Page,

University libraries also understand that they could save a significant amount of money, and better meet their mission of free access to information, by supporting open access. A consortium of universities called *Compact for Open Access Publishing Equity* (COAPE) is facilitating open access by, for example, contributing to OA journal publication fees for their faculty. Research Information Network (2008) estimates additional overall savings, on top of moving fully digital, by moving from a subscription-based to publication-based funding model (see also Houghton, 2009).

On April 17, 2012, Harvard University library issued a memo to its faculty titled "Major Periodical Subscriptions Cannot Be Sustained" (Faculty Advisory Council, 2012). It described how the subscription-based publishing model is an enormous financial drain and that the University can lead the way toward open access solutions. Among other things, it suggested that faculty "Consider submitting articles to open-access journals, or to ones that have reasonable, sustainable subscription costs; move prestige to open access" and "If on the editorial board of a journal involved, determine if it can be published as open access material, or independently from publishers that practice pricing described above. If not, consider resigning." This need not be a single university effort. For example, COAPE and the top 50 research universities could coordinate to establish an end-date, say three years ahead, for cancelling all journal subscriptions. In the intervening period, they could facilitate the transition of their faculties to publishing in OA journals, and reallocate subscription fees toward covering publication expenses. With a coordinated effort, the closed access system would decline rapidly.[8]

It is not surprising that many publishers are leery or actively resistant to open access. However, not all publishers are opposed to OA models. The Open Access Scholarly Publishers Association (OASPA; http://www.oaspa.org) represents the interests of dozens of publishers that are supportive of open access (e.g., SAGE publications, American Physical Society, Oxford University Press, BMJ group). These publishers are at the vanguard for facilitating the embrace of new technologies to expand the accessibility and exchange of scientific works.

**Barriers to change**. Remarkably, the group that is presently contributing the least to a move from closed to open access is scientists themselves. First, the costs of the present publishing system are opaque to scientists at highly-resourced institutions because they can access the articles they need and do not see the costs borne by the public or university. Second, publishers often send a small portion of their revenues to scientific societies to keep them invested in (dependent on?) the publishers.

Third, scientists hand over copyright free-of-charge, are paid very little for editorial services, and volunteer their time as reviewers to closed access journals. One estimate of the total value of volunteer peer review services was more than $3.0 billion globally (Research Information Network, 2008). If journals paid for the peer reviews, subscription prices would need to increase by 43% to cover the expense. Even so, publishers often claim peer review as one of their provided services justifying the high costs. For example, an APA presidential column about OA noted, "publishers add immense value through such functions as editorial selection, peer review, copyediting, and design production." (Brehm, 2007). For many closed access publishers, the present business model has been a remarkable success. For example, in 2011, Elsevier – the largest scientific journal publisher with about 2,000 titles – reported

[8] Access to archives would remain an expense, but of minor relative magnitude because of the elimination of production costs.

a profit of £768 million ($1.2 billion) on revenue of £2,058 million ($3.3 billion; 37.3% profit margin, up from 36% in 2010; Anon, 2011). Another large publisher, Springer, showed a similarly stunning 34% profit margin (Anon, 2011). Apple Computer's profit margin in 2011 was 24% (Taylor, 2012), just under publisher Taylor and Francis's 25% (http://www.informa.com/documents/INF2570%20AR10%20cover%20AW05.pdf).

Fourth, scientists still mostly publish in closed access journals and demand that their universities pay for subscriptions to those journals. The closed journals have developed strong brand identities. Those identities provide heuristic information about the prestige and topic of what is published in the journal. Switching to new journals that do not have that accumulated reputation is a risk, particularly for early career scientists who rely on the reputation building mechanisms of where they publish.

Most scientists are unaware of the implications of their choice to publish in closed journals. Awareness could increase scientists' preference for OA outlets. Further, a practical appeal to scientists is evidence suggesting that OA articles reach more readers (Davis et al., 2008), and have a citation impact advantage (Antelman, 2004; Eysenbach , 2006; Gargouri et al., 2010) over closed access articles. Even so, it is not reasonable to expect that scientists will easily stop publishing in the journals that they know and value. The scientific community could contribute to the change by establishing good reputations for OA journals. One approach is to realize that the brand value of a journal is *not* tied to the publisher – it is tied to the prior published work and the scientific community that supports and runs the journal. If, for example, the *Society for Experimental Social Psychology* (*SESP*) decided to end its relationship with Elsevier for the *Journal of Experimental Social Psychology* (*JESP*) and move elsewhere – which would the scientific community follow – Elsevier or SESP? When we ask this question of colleagues, it elicits a laugh because none identified the scientific brand with the publisher. A common response is "I did not even know who published *JESP*!"

Scientists may be more likely to change their own publishing behavior toward open access if they are assured that they will not be taking a career risk by doing so. The most straightforward way is to make the change collectively. Individuals, or societies, could self-organize collective action by gathering signatories for behavior change triggers: e.g., “I will make OA journals my submission destination of choice as soon as (a) 50 of the top 200 cited scientists, (b) 400 faculty from the top 50 research universities, or (c) 1000 academic faculty in my discipline also sign on to this commitment.” This lowers the risk of behavior change because the commitment is contingent on collective action. Another possibility for reducing risk is to abandon closed journals sequentially by collectively boycotting a single publisher. This is being pursued by the Cost of Knowledge boycott of Elsevier (http://thecostofknowledge.com/). Started January 22, 2012, by May 3, 2012, 11,081 scientists had joined the boycott of publishing, editorial work, or reviewing for Elsevier journals. If successful, other publishers would either change practices or be next in line for a boycott.

Journals can migrate to open access platforms based on the decision of the publisher changing its model, the society that owns the journal changing its publisher, or via the editorial board even if a society does not own the journal name. Most editorial boards are scientists motivated to serve the scientific community, and not the commercial goals of the publishers. If the owner of a journal does not want to switch to an open access format, editorial boards could move to a publisher that supports open access. They can add the word “Open” to the previous journal name and continue to operate as before, with the same reputation and review standards, now through an open access outlet.

There is a particular challenge for societies like the American Psychological Association (APA) and the American Chemical Society (ACS) that publish their own journals. Societies like ACS earn millions per year on subscriptions to its journals (Marris, 2005). One might hope that scientific societies would see their mission for the free exchange of information overriding their desire for an enhanced revenue stream. But, once established, revenue streams rapidly become indispensable. APA and ACS have both expressed some skepticism of open access approaches (APA, 2008; Brehm, 2007). ACS issued a position statement "Ensuring access to high quality science" (ACS, 2010) that opposed OA policies stating, among other thing, "Initiatives that mandate the open deposit of accepted manuscripts risk destabilizing subscription licensing revenues and undermining peer review." Also, ACS has hired public relations and lobbying groups to counter the movement toward open access (Bielo, 2007; Giles, 2007).

A final challenge is achieving a smooth adaptation of the funding model. Scientists are understandably leery when the financial model changes so that authors pay up front. The key insight is recognizing that it will be the same money (and less of it; Research Information Network, 2008; Houghton, 2009) just being moved from subscription access to publication costs. Universities and funding agencies will address the shift in the financial model through reallocation of subscription fees to publishing fees. Scientists just need to provide them with the leverage to shift the resources from closed access subscriptions to open access support.

A policy of fully OA digital delivery means that anyone can read, learn from, and critique the scientific literature from anywhere in the world with just an Internet connection. Institutions with smaller budgets would have the same access to the scientific literature as major research institutions. The scientific communities in emerging nations would have access to the existing literature in order to learn and develop their own knowledge infrastructure to make contributions to science. The thousands of PhDs that are working outside of academic institutions would still be able to access, consider, and apply the latest scientific knowledge to their professional responsibilities. For example, mental health professionals are more likely to read a scientific article if they have free access to it (Hardisty & Haaga, 2008). Additionally, the funding public would have the ability to see directly what their tax dollars are supporting. This is particularly relevant for research with policy implications. The more widely available the research literature, the more likely that the many minds will be inspired to pursue new discoveries, find problems with the existing claims, or create applications that do not yet exist.

**Stage 3: Disentangling publication from evaluation**

The major shift in Stage 2 is to dramatically increase accessibility. The major shift in Stage 3 is to accelerate communication by making the publication and evaluation of articles distinct steps in the scientific communication process. Authors prepare their manuscripts and decide themselves when it is published by submitting it to a repository. The repository manages copyediting and makes the articles available publicly. This changes the standard practice by altering the editor's role. In the standard practice, the roles of gatekeeper and evaluator are confounded in the journal editors and their *ad hoc* review teams. In the revised practice, the gatekeeping role is given to the authors; the editor's role is evaluation.

Scientists commonly work on problems that are similar or related to those investigated by other scientists. The pace of dissemination can influence the direction and maturation of a research discipline dramatically. One scientist's results might alter the direction or strategy of another scientist's research. The feedback between independent laboratories accelerates the accumulation of knowledge. Stage 1 accelerated dissemination by reducing the lag between acceptance of an article and its appearance "in print." Stage 3 eliminates the remainder of the lag.

This change is nearly complete in physics. In 1991, arXiv emerged as a mechanism for distributing preprints among a small group of physicists working on related problems. Today, operated by Cornell and supported by more than 50 universities, http://arXiv.org/ is a public repository of manuscripts for a large portion of physics research (and now math, computer science, and quantitative sciences).[9] Most physicists post their manuscripts to arXiv when they are completed, and use arXiv to keep up-to-date on new research in their area. ArXiv is organized into topical areas, and has a number of features for finding relevant research. Posting manuscripts on arXiv does not replace publication

[9] ArXiv is very cost-efficient too. The 2012 direct costs were $589,000 (http://arxiv.org/help/support/2012_budget) for a repository that serves an entire field.

officially – most scientists still submit and publish the manuscripts in peer-reviewed journals. But, functionally speaking, the arXiv repository makes peer review a secondary step. Once posted on arXiv, the research is disseminated to the scientific community and may start influencing other scientists (see http://ssrn.com/ and http://repec.org/ as two examples for social sciences and economics respectively). If the authors revise the manuscript, such as after receiving peer review at a traditional journal, updated versions are posted and a version history of the article is maintained.

Usage of manuscript sharing mechanisms has grown dramatically in the last 10 years. The *Registry of Open Access Repositories* (http://roar.eprints.org/) reported a growth of repositories from less than 200 in 2004 to more than 2200 in 2010. In 2012, *Google Scholar* began ranking journals based on a variation of the *h*-index for journal citation impact. Three self-archiving repositories (RePEc, arXiv, and SSRN) were among the top 10 journals in this index of impact (top 10 in order: Nature, New England Journal of Medicine, Science, *RePEc, arXiv*, The Lancet, *SSRN*, Cell, Proceedings of the National Academy of Sciences, Nature Genetics; http://scholar.google.com/citations?view_op=top_venues as of April 16, 2012).

Besides accelerating dissemination, separating publication from evaluation reduces the file-drawer effect (Rosenthal, 1979). If scientists take the time to write up and post a report, then it will be part of the record even if the authors give up on getting the report through the peer-review process. It also provides access to the evolution of a contribution. When the only public record is the final publication, other scientists do not have the opportunity to see the history of the article through the review process. Separating publication from evaluation makes that history available, should it be of interest to others. For example, editors and reviewers may require authors to drop methods or results that are of tangential interest or inconclusive, even if specialists in that area would gain from having access to that information for replication, extension, or learning the details of the research.

**Barriers to change**. There are two common concerns about separating publication from evaluation. First, science is competitive and authors worry about getting scooped – another laboratory publishes the same results first. In physics, this concern disappeared once authors realized that, because arXiv posting date is registered, the first laboratory to post the research on arXiv earned the "first finder" accolade. Peer review became certification of a job well done, not "who did it first."

The second concern is the fact that the peer review process provides an important filtering function. Ideally, peer review removes the poor research and retains the good research – i.e., it acts as quality control (Armstrong, 1997; Goldbeck-Wood, 1999). That filter is important for deciding how to spend one's limited time and energy for reading the research literature.[10] For Stage 3, this concern is immaterial. Physicists who do not want to spend time looking at articles in arXiv can wait until the articles appear in the "traditional" journals following peer review. For these scientists, nothing is different from the present standard practice. The key change is that scientists do not need to wait for peer review if they do not wish to do so.

On this point, we considered our own research practices. One of our core areas of expertise is implicit social cognition and implicit measurement (Nosek, Hawkins, & Frazier, 2011). There are many scientists in our research area that we follow and respect because of their established reputation for doing good research. We want to know what they have done as soon as it is available because it could affect our research. Also, we want to be aware of all research that is *directly* relevant to what we do, even if it comes from people that we do not know. For some articles, we might not read past the title or abstract; others might be immediately and obviously important for us. If the case study is a reasonable estimate of standard practices (see Table 1), a repository will give us access to this information two years sooner than the present lag time in dissemination.

---

[10] Outlets like arXiv provide "light" filtering services, not full peer review (for details see: http://arxiv.org/help/general).

For areas outside of our core interest and expertise, we can afford to wait. For example, we want to stay informed on emotion research, but we are not emotion researchers. Knowing the cutting-edge is less critical for our daily research effort. Waiting for peer review means that emotion experts are providing us advice about what is worth reading. For areas even more distant to our core expertise, we might delay even longer – reading only those articles that become citation classics well after initial peer review is over. The key result of Stage 3 is that specialists in an area can be aware of what is happening in related laboratories immediately. Further, all written reports, published or not, are available for review and meta-analysis, reducing the file-drawer effect. This might increase the likelihood that people get credit, at least in their own field, not only for significant results but also for rigorous and creative work that did not produce significant results.

**Stage 4: A grading evaluation system and a diversified dissemination system**

Through Stage 3, nothing in the scientist's daily practices, save for the editor's role, must change. The changes are, instead, in the communication ecosystem. Scientists can continue operating exactly as they do presently; the changes just increase opportunity for more and faster access to others' research. Stage 4 is the first stage in which scientific communication necessarily changes for the scientist. In this stage, we alter the peer review system and diversify the dissemination system.

The standard practice has an article evaluated for the journal considering it, and then publishing articles in one and only one journal. In Stage 4, we separate the link between peer review and specific journals. Instead of submitting a manuscript for review by a particular journal with a particular level of prestige, authors submit to a review service for peer review (see, for example, RIOJA, http://www.ucl.ac.uk/ls/rioja/about/, designed as a peer review overlay to arXiv). The review service does not decide whether the article meets the "accept" standard for any particular journal. Instead, it gives the article a grade. Journals could retain their own review process, as they do now, or they could drop their internal review system and use the results of one or many review services. Because all articles are published (Stage 3), journals are not publishing articles, they are promoting them. Journals would have no exclusivity claim on individual articles. Dozens of journals could promote the same article. For example, SSRN (http://SSRN.com/) has many research networks and eJournals for filtering and disseminating the >300,000 papers contained in their repository. Authors can submit their paper to as many eJournals as they like and editors consider its relevance for dissemination through their portal.

This change splits the editor role further. Most of the editorial infrastructure of journals gets consolidated into review services (e.g., APA journal boards could consolidate into a single review service for all of psychology), and a distinct, "curator" editorial role emerges that is just for selecting articles to be promoted in any given journal.

**Filters for quality and importance of research**. One function of the present system of peer review is quality control (Armstrong, 1997; Goldbeck-Wood, 1999; Horrobin, 1990). An expert panel provides an evaluation, not just of the science in its own right, but as a guide to potential readers. Reviewers evaluate whether the research meets the importance and quality standards of the journal considering it. Editors, reviewers, and authors recognize that journals differ in prestige, and calibrate their evaluations accordingly. This evaluation process is essentially a grading system of scientific articles with the journal's prestige providing the grade. Scientists do not have time to read everything. Journal prestige provides a useful, simplifying heuristic for managing overwhelming information. Presently, if an article does not meet the journal threshold for quality or importance, it is rejected and submitted elsewhere.[11] The resubmission engages a new set of editors and reviewers. The process recurs until a review team agrees that the article meets the journal's threshold. In the case study, for example, 49% of articles were reviewed at more than one journal.

---

[11] Evaluating "importance" is a challenging prospective task. *PLoS ONE's* dismisses "importance" as an evaluation criterion in their mission*:* "PLoS ONE will rigorously peer-review your submissions and publish all papers that are judged to be technically sound. Judgments about the importance of any particular paper are then made after publication by the readership (who are the most qualified to determine what is of interest to them).*"* *(*http://www.plosone.org/static/information.action). In 2006, its first year of existence, *PLoS ONE* published 138 articles. In 2011, PLoS ONE published 13,798 articles. It is the largest journal in the world. Despite explicitly rejecting "importance" as an evaluation criterion, PLoS ONE has an impact factor of 4.4. That puts it in the top 25% of impact factors for biological sciences.

In Stage 4, the task of the editor and reviewers is to grade the manuscript, not to decide if it should be published or not. The manuscript is already published (Stage 3). Authors submit manuscripts to review services that are not connected to any particular journal. Authors could choose to submit to a review service before or after they publish the article at the open repository. The editor, with guidance from the reviewers, gives the article a grade, along with recommendations for how the paper could receive a better grade. The authors have a number of options: they can settle on that grade, revise the paper and resubmit to the same editor, or revise the paper and submit to a different editor. When readers view a published article in the open repository, they could see its current grades (potentially from multiple review services), and also go back to the history of the article, viewing its previous versions and grades.

The major benefit of this change is to shorten the review cycle or, more accurately, increase the authors' control of the length of the review cycle. In the present system, authors tend to submit to the most prestigious outlet that they believe might accept the article and then continue down the prestige rankings until it is accepted. This amounts to an evaluation system of gradual exclusion – it isn't an A, submit again; it isn't an A-, submit again, it isn't a B+, submit again; it is a B, accept, done. If the authors had received the grade on the first submission, they may have recognized and agreed with the limitations identified by reviewers, or they might have disagreed and conducted a revision focused on addressing those shortcomings to improve the grade. In the present system, it is easy to keep sending rejected articles to new journals – little effort is required by the authors once the paper is written. In the new system, every decision letter is a "revise and resubmit," as all articles are already "accepted." Resubmissions require direct effort to address the concerns, and the evaluation standards stay constant because reviewers are not calibrating based on journal prestige. Authors would likely be more likely to weight their chances for improving their grade against the effort required to do so, so that they are focusing their efforts on manuscripts that are most likely to benefit from revision.

Another benefit of this change is that it would shift the emphasis of peer review. Presently, a prominent factor is whether the article meets “this journal’s” standard. Editors must make judgment calls on mixed reviews for whether the authors should get another chance to meet the standard, or be rejected outright and move on. Here, final acceptance is the decision of the author, not the editor. Peer review spends no time evaluating whether it belongs, and instead focuses on what can be learned from the research, its limitations, and how it could be improved.

Because peer review is independent of dissemination, the peer review service would be independent of journals. Peer review could be managed by a single service or competing services with the armies of editors and reviewers that are presently spread out across thousands of journals. Different review services may offer different technological solutions (e.g., the website), different editor and reviewer pools, and different general guidelines and philosophy for reviews. Presently, most journals use similar review guidelines and are only different in subject area and type of papers. In the new system, authors could select to submit their paper to one or more review services based on the topic, methodology, and type of report. The review services could compete on prestige, by improving the quality of their grading system, and by offering unique approaches to appeal to authors based on topic, content or style of their reports.

In this system, authors would submit their paper for grading to the review service as many times as they wish. Because the version history and grades will be publicly accessible, authors will have incentives to address critical comments as effectively as possible and to avoid resubmitting many times. Submitting the same manuscript over and over again until the most recent grade happens to be high would be possible, but the behavior would be available publicly and thus have consequences for reputation.

**Filters for topic and content**. In addition to presumed quality or importance, journals today have identities– the type of science they publish – such as by specializing in different topical areas, short reports versus comprehensive investigations, review articles versus empirical investigations. These identities are important for sorting so that scientists can find research that is useful to them. With the separation of review from publication and dissemination, is this lost? No, in fact, digital and open access provide dramatic enhancement of opportunities for content filtering.

In Stage 4, the role of the editor is split in two – one editor for managing peer review, and a different editor for selecting which articles appear in a journal or other type of collection. The journals as they are presently known could continue to exist, but they would not need to conduct their own peer review (unless they wished to do so as another, specialized review service). Editors would either seek out content on their own from the results of the review services, or consider submissions from authors or others to be featured in the journal. Thus, the journals could draw from the same large pool of articles evaluated by review services for possible promotion in their journal. This evolution of the journal just barely scratches the surface of the possibilities when evaluation is separated from dissemination. In the present system, an article is published in one and only one journal. That made sense with paper, but there is no reason for it in the digital age. With articles available digitally, there can be an infinite number of filters for disseminating the content.

Many of these alternate filters exist already. For example, the current authors receive (a) weekly emails from the *Association for Psychological Science* highlighting articles from that month's journals, (b) citation alerts from *Web of Science* and *Google Scholar* whenever a new article is published that cites articles that are important to our research – regardless of the originating journal, (c) RSS feeds from science blogs that cover science on personally relevant topical domains, (d) new article alerts from journals, (e) a daily email from National Affairs blogger Kevin Lewis with citations and abstracts for a dozen or so articles culled and organized from hundreds of sources into a "special issues" about social science (e.g., one day is "racism" the next is "work-life balance," the next is "labor markets"), and (f) emails from colleagues in the same research area sharing new manuscripts. These expose the authors to a much wider array of the research literature than they receive via subscriptions to the key journals in their subfields. In an open system, all of these mechanisms of dissemination function as content filters.

Societies could maintain journals on topical areas relevant to their membership. Universities and departments could maintain their own journals. Interest groups could maintain journals to highlight, for example, research by young scientists, research from institutions that have relatively little research infrastructure, or research using particular samples or instrumentation. Labs and individuals could maintain their own journals. All would draw from the universe of published articles using whatever criteria they deem relevant. The same article can appear in dozens or hundreds of journals. Authors with a report that interests multiple disciplines would not need to make the difficult decision of "which discipline" by choosing a single journal for publication. Much like news and entertainment media filters (CNN, Huffington Post, Perez Hilton, Deadspin, Gawker, and the thousands of blogs and other filters), readers would subscribe to filters that are likely to inspire new ideas or alert the researcher to findings relevant to their expertise.

**Barriers to change.** One challenge is that individual scientists may get a biased exposure to the literature if they self-select narrow filters. For example, the diversification of news media makes it possible to live in a universe that only reports news consistent with how you already think (Sunstein, 2009). This may be occurring in the present scientific system to an even greater degree with the fragmentation of journals across disciplines. Even so, it is an important issue to address because innovation can be spurred with exposure to new ideas from unexpected sources. Professional societies might play an important role in having journal feeds going to all of their members that ensure dissemination of a diversity of topics that scientists may not have self-selected to receive.

Just as the digital age has introduced enormous challenges for the news media's business model, eliminating journals as we know them today will further decrease possible profit for publishers focused on the journal as their asset rather than the publishing infrastructure. If a group of volunteer scientists can publish their own journal that is as good as a commercial publishers', then there is little chance that the commercial publisher will be able to produce a viable business model around the journal. Most likely, publishers' role would be focused on infrastructure and support services for the repository, copyediting, and the review services. Some publishers might die, but the best will adapt and provide services that improve access and filtering of the content. Their profit incentives will align with science's incentives for free and easy access to information.

In Stage 4, the simple heuristic of "which journal published it" is no longer available. Evaluation for hiring and promotion can consider the grades of articles, and also which and how many journals disseminated the reports. Changing the rules of the trade make it more difficult to compare past and new scientists. All of this suggests that there will be some work to learn how to evaluate scientists in the new system. However, we consider this an opportunity, not a cost. The present system of evaluation – relying mostly on journal impact factors – is crude and fraught with error (Adler & Harzing, 2009; Cameron, 2005; Holden et al., 2006; Seglen, 1994, 1997; Starbuck, 2005), not to mention the fact that it can be influenced by self-fulfilling prophecies – the presumption that if a paper appeared in a more prestigious venue it is more worthy of reading and citing than the same paper in a less prestigious one. Even if the number of citations is a reasonable measure of impact for individual articles, a journal's impact factor is a weak predictor of the number of citations of a single article published in the journal (Holden et al., 2006; Seglen, 1994). In social psychology, for example, publishing in *JPSP* (IF = 5.2) is a major boost to earning an academic faculty position, whereas publishing in the *British Journal of Social Psychology* (*BJSP*, IF = 2.1) is solid, but certainly not a career-defining event. *JPSP* publishes about twice as many papers as *BJSP* each year. We examined citations of articles from the 2010 and 2011 issues of both journals and found that (a) between 35-40% of the total citations are to the top 10% of the published articles, (b) the bottom half of articles account for just 10-15% of total citations,[12] and (c) the bottom half of *JPSP* is essentially equivalent to *BJSP* in total citations. In sum, by citation count, *BJSP* and the bottom half of *JPSP* are the same, but the difference in career impact of being in one group or the other is dramatic. Further, journal impact factors only really reflect the small minority of high-impact papers in each journal and ignore the huge variability within journals.

Every year that goes by with journal impact factors as criteria to judge scientists is another year in the dark ages of scientific evaluation. Our observation is that most scientists agree with this, but simultaneously feel powerless to change it and thus still use journal impact factor for evaluation "because everyone else does." Breaking the constraint of each article being in one and only one journal will focus evaluation of quality and impact where it should be – on the article itself.

**Stage 5: Publishing peer review**

Peer review is a central scientific practice (Peters & Ceci, 1982). Confidence in the quality of design, analysis and interpretation is improved by the independent evaluation of expert peers. While ultimate confidence is in the reproducibility of the results, and by the eventual impact of the research and theory on other science, peer review presently serves as a quality control barrier to entry (Armstrong, 1997, Goldbeck-Wood, 1999, Horrobin, 1990). An editor and one to five reviewers decide whether the scientific community should see, read and be influenced by the work – at least in their journal.

In addition to the gatekeeping (removed in Stage 3) and evaluation (changed to a grading system in Stage 4), excellent peer review can identify confounds or problems in the research design, point out alternative analysis strategies that avoid inferential challenges, suggest new avenues of investigation that can clarify the validity or applicability of the hypothesis, or provide alternative theoretical understandings of the same empirical evidence. In other words, peer review can contribute to scientific progress. However, present practices do not take full advantage or give recognition to these contributions of peer review. Stage 5 corrects this by publishing peer reviews (see also Benos et al., 2006; Wicherts, Kievit, Bakker, & Borsboom, 2012). Reviewers conduct reviews for review services. Reviewers then decide if their reviews will be published in the repository alongside the originating article. If authors revise and resubmit the article, new reviews are attached to the resubmitted version. Readers have access to the evolution of the article and the reviews from each stage.

---

[12] These distributions of citations among top and bottom articles hold across the four other social psychology journals that we examined: Social and Personality Psychological Science (SPPS), Journal of Experimental Social Psychology (JESP), European Journal of Personality (EJP), and Personality and Social Psychology Bulletin (PSPB).

In present practice, there are few incentives to be a reviewer and to do a good job reviewing. Peer review is voluntary and usually anonymous. Peer review takes time. The most that it can do for reputation building is add a minor vita entry. Further, when reviewers do invest time into the process and provide an excellent, insightful review, the only knowledge gain from that effort is for the authors, editor and other reviewers. Only a portion of that scholarship influences the manuscript, and the reviewer gets no identifiable credit for the contribution. As frequent reviewers, we have observed reviews by others that provided us with insight and ideas that would surely have benefitted others. Besides being a loss of scientific contribution and a disincentive for doing a good job, the closed nature of the peer review process violates the scientific values of openness and transparency.

Stage 5 increases the incentives for high-quality peer review. Having given up lucrative careers doing something else, the scientist's primary currency is reputation. Scientists build reputation by contributing to science, primarily through publication. Stage 5 creates a new category of scientific contribution by giving peer reviewers the opportunity to publish their reviews. If reviewers choose to do so, the reviews are published under their name and are linked to the version of the report that they reviewed. The original authors have the opportunity to address those concerns whether they publish a revision or not. Reviews become a public scientific contribution, and scientists can gain reputation by being good reviewers. This already occurs in mathematics. The majority of math articles have published reviews that appear in *Mathematical Reviews* (http://www.ams.org/mr-database/) that identify the reviewer and are a basis of reputation building. Likewise, the journal *Biology Direct* (http://www.biology-direct.com/ ) publishes reviews and the authors' responses to the reviews (e.g., http://www.biology-direct.com/content/7/1/4).

Publishing reviews would increase transparency, and would make anonymous reviews infrequent rather than the norm. It would also be much easier to learn about the strengths and limitations of articles. Reviews often raise interesting limitations and open questions but recommend publication nonetheless. Because the critical points are public, authors would care more about addressing critical points, rather than focusing on doing just enough to convince the editor. Finally, a new class of contributor would emerge – scientists that rarely do their own novel research, but frequently offer critique and review of others. These contributors already exist, but they are unrecognized because the present system does not acknowledge or reward their contribution. This is underused potential, particularly considering that most trained scientists are not at high-output research universities generating research. But, they are trained, knowledgeable, and can offer great insight on what is produced by others. Further, it is likely that high-quality critiques would be cited by later articles either to raise or address the critique. It is easy to conceive of a future in which some scientists could earn tenure by being renowned evaluators of research rather than producing research on their own.

**Barriers to change**. Why do the present privacy norms exist? One rationale is the belief that reviewers will avoid being critical if peer review is public. However, there is little demonstrated evidence that public science is non-critical. Science is defined, in part, as skeptical inquiry. Casual observation suggests that public debate in science is active and frequent. Further, a randomized trial of blind versus open review found no difference in the rated review quality or the ultimate publication recommendations (van Rooyen, et al., 1999). Debates are active, frequent, and substantive. Further, reputation is established in science through effective critique and alternate perspectives. There is strong motivation for scientists to provide effective critique rather than be cheerleaders for each other. Also, when peer review is public, reviewers can gain – and lose – reputation via their reviews. Overly positive reviewers – and *quid pro quo* positive reviewing among friends – are easy to identity and undermine the credibility and reputation of the reviewers. The same is obviously true of overly negative reviews. A good review identifies the strengths, weaknesses, and opportunities of the research to foster knowledge, and does so constructively.

A variation of this concern is that the closed review system is particularly important to protect junior people from potential retribution when criticizing senior colleagues. Indeed, because of that concern – for Stage 5 – we did not recommend making transparency a rule, only a rewarded norm. However, we suspect that this concern is ill-founded and would fade rapidly. In part, the concern is paternalistic. Scientists make their reputation in the public sphere by elaborating or challenging existing ideas. More often than not, those ideas are from people more senior. Further, transparency increases, not decreases, accountability and ethical behavior (Lerner & Tetlock, 1999). Transparent peer review makes it easier to detect when a senior person is trying to seek vengeance on a more junior person. In the existing system, senior scientists can use their position and power, rather than logic and evidence, to affect outcomes privately. Senior scientists can torpedo an ego-challenging manuscript with a hostile review, and almost no one is the wiser.

Transparency makes the positions, claims, evidence, and style of reviews available for all to evaluate (Smith, 1999). Authors and reviewers can gain or lose reputation based on the quality of evidence and critique. And, transparency increases accountability for offering the critique in a professional manner. If nothing else, transparency's effect on tone could transform hostile attacks into cogent critiques. Further, when information about reviews and grades is accessible freely, research can be performed to detect biases and heuristics that influence reviews, and may educate the scientific community on how to overcome these influences (Peters & Ceci, 1982; Petty et al., 1999; Reich et al., 2007; Wilson et al., 1993).

**Stage 6: Open, continuous peer review**

As a gatekeeping function, peer review relies on the expertise and attention of a few judges. That attention is time-limited, in that once the editor issues a decision, peer review is complete. But, of course, that is not really the case. Most peer review occurs informally after publication among scientists who are reading, evaluating, critiquing, and applying the published research. This evaluation can also evolve over time. An exciting demonstration might get published in a high prestige outlet, but enthusiasm will dissipate rapidly if a critical confound is identified. Another article might have substantial difficulty getting published, but may come to be appreciated as an effective challenge to prevailing wisdom over time. Except for the rapidly growing community of science bloggers, almost all of the dynamic post-publication discussion appears in unpublished conversations in labs, reading groups, and between individual pairs of colleagues.

Stage 6 opens the review process so that all members of the community can contribute and evaluation can evolve over time (Arms, 2002; Harnad, 1998). The formal, editor-based review process, now managed through review services, becomes just one component of evaluation rather than the only evaluation. Reviews from the solicited review services are posted in the repository next to the article (Stage 5), and a commenting system is linked to each article with reviews from unsolicited review services and from single reviewers. Readers can comment, ask questions, and grade articles; authors can reply; all can discuss. Comments are evaluated – reviewing the reviewers – with positive and negative votes. Grades of articles and comments are aggregated to provide summary statements of the article and reviewer points. Open reviewers accumulate reputation status by the comments they make to articles.

Even though this is the last stage of our publishing and review utopia, there are already a variety of journals that are developing these practices. *PLoS ONE*, for example, has an open commentary system linked to each article (see, e.g., http://www.ploscollections.org/article/info%3Adoi%2F10.1371%2Fjournal.pcbi.1000037). The *Journal of Medical Internet Research* has added an open peer review option to its standard review process (http://www.jmir.org/reviewer/openReview/abstracts). One of the most interesting examples is *F1000* (http://f1000.com/), a new journal launching in 2012 for biology and medicine, is a fully open-access journal that provides immediate publication, open peer review, open data, and flexibility to update published articles with new versions.

A major limitation of the present review system is that it depends on a small number of experts to identify the strengths and weaknesses of a contribution. Even by publishing those reviews (Stage 5), it is unlikely that those reviewers will always have sufficient expertise to evaluate the theory, design, instrumentation, analysis strategy, and interpretation of every component of the article. In an open peer review system, many more minds can contribute their unique expertise to the evaluation of an article. For example, in the last month we read two published articles in high-profile outlets that had critical errors in the analysis of the Implicit Association Test (IAT; Greenwald et al., 1998), a technique for which we have specialized experience and expertise. It is likely that the selection of reviewers did not include an expert on this measurement technique. In an open review system, we could have taken five minutes after reading the article to post a short comment to point out the error and recommend a fix. The authors could, if they desired, reanalyze the data and post an updated version of the article. By contrast, in the existing system, the only mechanism for addressing such errors is to write and submit a comment to the publishing journal, and perhaps conduct and ultimately publish a reanalysis. In that case, besides the enormous lag time, the original articles would persist and other researchers might erroneously reuse the original, erroneous analysis strategy. Indeed, a colleague recently emailed us noting that an editor recommended an analysis for her project citing one of these original papers. It is very difficult to get such an error out of the present system.

Scientists who do not have the resources or interest in doing original research themselves can make substantial contributions to science by reviewing, rather than waiting to be asked to review. Crowdsourcing has demonstrated enormous potential for evaluation and for problem solving (see Nielsen, 2012). Companies such as Amazon (http://amazon.com/), cNet (http://cnet.com/), and Yelp (http://yelp.com/) use the collective wisdom of volunteer reviewers to recommend books, restaurants, electronics, and everything else to other consumers. Discussion portals such as Reddit (http://reddit.com/) and Slashdot (http://slashdot.org/) use the crowd to promote stories of interest to their communities and to evaluate commentary about those stories – positively evaluated commentary is made more prominent and negatively evaluated commentary is made less visible. Amazon, for example, offers simple mechanisms for reviewing the quality of reviewers (http://www.amazon.com/review/top-reviewers). These ratings are accumulated and top reviewers are highlighted with labels such as "Top 500 Reviewer" on their reviews. Their earned reputation confers status and attention to their subsequent reviews.

Switching to an open model of reviewing has a number of additional benefits besides incentivizing reviewing. First, those that are not interested or skilled at reviewing do not have to do so. The present system employs a "somewhat true" assumption – that the producers of knowledge are the most expert to evaluate others' production of knowledge. Certainly the expertise in doing the research is important, particularly for evaluating methodology. However, producing and evaluating employ distinct skill sets. Some producers are terrible reviewers, and many potentially effective reviewers are underused because they are not frequent producers. For example, there is enormous underused expertise outside of research universities among very smart, very skilled scientists who are at institutions with a stronger teaching than research emphasis. Many are asked to review in the present system, but some might embrace a more prominent role in reviewing as their primary scholarship activity if given the opportunity to do so.

Second, open review does not rely solely on the wisdom and expertise of the editors and selected reviewers for judgment of the entire contribution. Crowdsourcing is more likely to ensure that people with the right expertise have opportunity to weigh in and contribute to the evaluation. As in the example above, our expertise with the IAT can be applied across the dozens of articles of articles we read each month with bite-sized reviews on its design and analysis, leaving review of the rest of the article to others. The time investment would be equivalent to full review of just a couple of articles – especially considering that these are articles we are reading anyway to keep up on our field. Kaggle (http://www.kaggle.com/) has created a business model based on this idea, and http://www.hypothes.is/ is attempting to apply the open peer review concept to evaluation of everything on the Internet. Present problems to the crowd and someone with the relevant expertise will see and be able to solve it much more rapidly than the originating person or organization would have been able to solve it themselves.

Finally, changing to an open review system would have a radical effect on the role of peer review for authors. In the existing system, peer review is a barrier to the authors' objective – publishing the article. Authors submit articles to journals that they think should accept the article. Reviewers usually prevent that from occurring. However, since Stage 3, the scientist's ultimate objective is no longer to get published, because everything is published. The objective is to influence future ideas and investigations, i.e., what should be the key incentive in the first place (Nosek, 2012). With openness, peer review becomes an asset to the authors. Work that is disinteresting will not be reviewed. Work that is interesting will get reviewed a lot. Reviews become the life blood of evaluating, improving and making the research have impact. The biggest threat in an open model is not to be reviewed, it is to be ignored.

**Barriers to change**. The biggest challenge in existing open review systems is the existence of "trolls," which is the Internet word used to describe people who make negative comments just in the hopes of causing agitation in others. In open systems, trolls post inflammatory, inaccurate, or extraneous comments disrupting normal operation and discussion. Certainly this would be an issue for open peer review, particularly for topics that generate widespread interest and controversy. There are effective solutions, the most critical of which is transparency. With anonymity, trolling is easy. Requiring confirmation of one's actual identity in an open review system means that misbehavior will impact the commenters' most valuable possession – reputation. With evaluation of reviewers, a troll accumulates a negative rating and future commentary is ignored. There are even automated mechanisms to "clean" commentary so that highly rated comments and reviewers are easily viewable and low-rated comments and reviewers are only seen if the reader deliberately asks to see all comments (see, e.g., http://reddit.com/ commentary system).

If a review system were completely open to the public, then even transparency would not be effective for individuals who do not care about their scientific reputation. For example, research on politically sensitive issues could be overrun by ideologues who care little for the scientific commitment to following the evidence, wherever it may lead. As such, a reasonable restriction to participation in open peer review would be membership in a professional society. Also, the editor-driven evaluation system would still exist. It would operate like cNet (http://cnet.com/), Rotten Tomatoes (http://rottentomatoes.com/), and other reviewing organizations that have both internal/expert and open review systems presented side-by-side. A discrepancy between these two evaluation modes offers an opportunity to identify biases in one or both approaches.

A final concern is that allowing evaluation past the initial review means more work for the original authors. After spending their time dealing with the original editor-based reviews, they feel done with the article. We believe that researchers will know how to allocate their time best. Some will find it important and useful to continue discussing their work with the scientific community even after an esteemed review service graded it. Others will prefer to shift their efforts elsewhere and allow the community to discuss the article on their own.

A common present day mindset, particularly among young scholars, is that publication signals the completion of a project. In practice, however, publication is only the beginning. The contribution only matters if it is read and influences others. Open evaluation is already occurring on a daily basis – the authors are just not exposed to it, and have little opportunity to respond. Establishing evaluation as an open, community-based system provides the authors an opportunity to hear, elaborate, and address the commentary. Further, receiving comments is another indicator of impact. In present practice, such commentary happens over the course of years with challenges and responses occurring in the published literature. Here, commentary occurs in real time. The key substantive issues are raised and discussed rapidly. If the discussion becomes repetitive or unresolvable, it dies out. The open system allows the same process that is happening in science across months and years to occur on a much shorter time scale.

**The New Reality**

Hari Seldon, psychologist of the future, spends the first 20 minutes of each workday morning browsing new articles. This morning he received the weekly journals from the *Association for Psychological Science* and his favorite for getting exposed to new work and new people –*Psychology by the New Scientist*. That journal is operated by a consortium of faculty and grad students from the University of California schools. The group believes that it is important to promote young scholars. So, they review and select high-quality articles (90+ ratings; 0-100 rating scale) that are first-authored by early career graduate students, a *JPSP* of early career scholarship. Hari also had a few new articles in his inbox from his evaluation filters: (1) *95+* articles in any psychology or closely related discipline and *85+* articles from his subdiscipline from his two favorite review services, (2) articles that received *90+* ratings by the reviewers he follows, (3) articles from his discipline that hit 100 comments with an average grade of 80+, (4) articles from his field that hit 100 times cited, or (5) articles from his broader interest fields that hit 1000 times cited. He also had a half dozen articles that were published yesterday from his content filters: (1) anything that cites one of the two most important review articles in his field, (2) anything that cites his own most important articles, (3) anything by the few dozen researchers in his subdiscipline that he follows, and (4) articles that use one of many keywords that are relevant to his research. He skims the titles and abstracts and marks four of the articles as worth looking at more closely later in the week.

Next, Seldon opens his comments box. He follows comments on his own articles and those of a number of other articles that are particularly relevant or interesting to him. There are three important comments to look over, two on his own papers. One is a mostly positive review by a highly-rated reviewer of a paper he published three months ago. Finally! The review acknowledges the importance of the work and the high quality of the methods in the reported studies, but raises an alternative explanation that he had not considered. Seldon recognizes that the argument may be worth addressing. As he starts to write a note to the first author, his graduate student, he laughs. The student has already written to him with a plan for responding as a comment. Seldon replies "Looks good, but don't waste too much time on a comment yet. The article is getting more attention now. We should let the discussion continue for a while without us. If we respond too quickly, we might short-circuit the debate and ideas that come up from others in the community. Once the comments have settled, we can do a revision to address the key critical points from the debate. The article has some good grades, you should be proud. Remember, the criticism helps us hone our argument. It's a good thing!"

The second comment is by someone he does not know about an article that he published just three days earlier. The commenter points out a small but important problem with an analysis technique that Seldon was using for the first time. The commenter is on-line, so Seldon immediately asks her a couple of follow-up questions, and she responds immediately. The fix is straightforward. He thanks the commenter and plans to re-run the analysis and post the fix later that day.

The last comment is from his “nemesis.” They are having an intense debate about the implications of a recent article that neither of them had written. He shakes her head at the silliness of the new argument, but resists firing off a snarky reply. He noticed that the debate is being followed by more than 100 others, including the original authors, so he wants to make sure that he can think through the argument and spend time editing for professionalism. Each of his comments so far has received a lot of positive evaluations. This could wind up being one of his more important contributions!

Seldon spends the rest of the morning working on a paper that is close to completion. He decides that it is time to publish. Hari prefers to publish and accumulate some open commentary for a few months for a revision before submitting it to a review service. His research area has a number of excellent reviewers that could make a real difference in the article quality prior to getting the certified reviews. After the last edits, he sends an “Okay to publish” message to collaborators. Hopefully, it will generate some interest.

After lunch, the faculty meets to discuss the two finalists for their open faculty position. The candidates are in the same research area, and today’s discussion revolves around the quality versus quantity dilemma. One of the candidates has published 19 manuscripts during her time in graduate school, about double the average applicant in her stage. Five of them received review service grades above 85. The rest had lower grades or were not evaluated at all, attracting only a few open comments. Five grades above 85 is certainly a strong achievement. The second finalist has just 5 published articles. However, the second candidate has a better weighted score on all three of the more popular grade weighting metrics (APS-certified, the Frazier-Lai technique, and the straight average). Advocates of the second candidate suggest that the first candidate sometimes publishes before a project is ready, and note that two of the second candidate's articles had a 98 grade from their review services. 98! Many faculty have never published even one article with a grade that high. Advocates for the first candidate counter that the open commentary evaluations are "only" 91 for one of these two papers, and that the other is not yet getting much open evaluation from the field. But they concede that there is apparent talent, just not as much data to go on and definitely less productivity.

Committee members agree that the possible tie-breaker is the quality of the candidates' reviews. One candidate's review ratings were average. The other candidate published a number of highly-rated reviews that were acknowledged as influential for article revisions, and are even being cited regularly in new articles. The candidate demonstrated great analytic talent and vast knowledge in a number of areas that do not appear in her own publication record. The faculty agrees to go back and do a more thorough qualitative evaluation of the top three articles and reviews from each before making a final decision.

Seldon returns to his office and is confronted by his newest student who looks devastated. He found an unrated manuscript that was published only an hour earlier that looks very similar to the study they were designing as his master's thesis. They sit down together and read over the design. It is similar, but Hari can be genuinely reassuring. "This is actually very useful. They did do the first part of what we are pursuing, but not the second part. And, look, their manipulation is very clever. We can use it with a couple of simple changes to pursue our second question. And, now we have some confidence that the manipulation will be effective!"

Finally, he gets her opportunity to prepare a response for her on-going debate, and then fix the analysis problem with her most recent publication. If his students don't keep interrupting perhaps he can do both quickly and then spend the rest of the afternoon writing that chapter that he has been thinking about all week.

## Conclusion

The technical and financial challenges of moving toward our scientific utopia of scientific communication are solvable. The major question for the reader is whether this is a utopia worth pursuing. If it is, how can we get closer to it? We selected a series of stages to imply an implementation plan for a broad scale move toward open communication. Also, each stage has existing small-scale examples providing evidence that the ideas are viable, and could be enhanced further without waiting for completion of prior stages.

There are reasonable alternatives to these suggestions worth considering, and specific implementation challenges, particularly coordination among interest groups, to translate ideal model to actual practice. However, the power to change rests with scientists first. Scientists and societies can support their library's effort to move from subscription-based to publishing-based funding models for open access. Scientists can submit to open access journals and participate in post-publication peer review in web forums. Most simply, scientists can talk about the scientific process that we want, rather than just accept the one that we have. We may never reach utopia, but we can improve on the present reality.

If the scientific community manages to accomplish the changes suggested in this article, we believe that daily laboratory practice would be largely unchanged, but that the flow of information would be much greater. Scientific communication would start early in the research process instead of only beginning after the date of publication. New results would be distributed widely and instantly. Limitations of research would be identified and addressed more quickly. Finally, scientists would spend more time thinking about the implications of others' results, and pursuing new lines to test them, rather than trying to decide if their results should appear in print or not. In the case study, an average of 677 days elapses between the initial submission of an article and its publication. We believe that scientific progress would be much further along after those 677 days if publication occurred on the first day, and the rest of the time was spent on critique, revision, and follow-up by the community-at-large.

There are many other components of the scientific process that could benefit from openness or other reforms to better align scientific values with scientific practices – e.g., open data (Reichman, Jones, & Schildhauer, 2011; Wicherts, 2011; Wicherts, Borsboom, & Molenaar, 2006; Yarkoni et al., 2010) and open workflow (Mathieu et al., 2009; Morin et al., 2012; Nosek, 2012; Schooler, 2011; Stodden, 2011; http://openscienceframework.org/). Science will benefit if the scientists collectively look up from the bench once in a while to evaluate and improve their daily practices and disciplinary norms. Together, we can maximize the quality of our systems to make more rapid progress in building a cumulative knowledge base of nature.

Table 1. Submission and publication history of all 62 unsolicited articles co-authored by Brian Nosek (July 1999 - April 2012)

Notes: Column IF = Impact Factor (U = Unknown), O = Outcome (A = Accept, R = Reject, U = Unknown), Att = Number of journals attempted; Article IF = average number of citations per year for the two years following publication in ISI database, the same database used for calculating journal IF (blank if not available or less than two years since publication); Journal abbreviations: AERJ = American Educational Research Journal; ASAP = Analyses of Social Issues and Public Policy; BASP = Basic and Applied Social Psychology; BRAT = Behaviour Research and Therapy; BJSP = British Journal of Social Psychology; CE = Cognition and Emotion; DAD = Drug and Alcohol Dependence; EP = Experimental Psychology; ERSP = European Review of Social Psychology; GD = Group Dynamics; GPIR = Group Processes and Intergroup Relations; HR = Human Reproduction; IJMH = International Journal of Men's Health; JAbP = Journal of Abnormal Psychology; JApP = Journal of Applied Psychology; JASP = Journal of Applied Social Psychology; JCR = Journal of Consumer Research; JEP = Journal of Educational Psychology; JEP:G = Journal of Experimental Psychology: General; JEP:LMC = Journal of Experimental Psychology: Learning, Memory, and Cognition; JESP = Journal of Experimental Social Psychology; JFP = Journal of Family Psychology; JHCPU = Journal of Health Care for the Poor and Underserved; JNMD = Journal of Nervous and Mental Disease; JPSP = Journal of Personality and Social Psychology; JSCP = Journal of Social and Clinical Psychology; JSPR = Journal of Social and Personal Relationships; JSSR = Journal for the Scientific Study of Religion; PA = Psychology and Aging; PS = Psychological Science; PSPB = Personality and Social Psychology Bulletin; QJEP = Quarterly Journal of Experimetal Psychology; SPPS = Social and Personality Psychological Science; SJR = Social Justice Research; SRA = Stigma Research and Action; SM = Statistical Methodology; SSM = Social Science and Medicine

| | | 1st Submission | | | 2nd Submission | | | 3rd Submission | | | 4th Submission | | | 5th Submission | | | 6th Submission | | | | Outcomes | | | | |
|---|---|---|---|---|---|---|---|---|---|---|---|---|---|---|---|---|---|---|---|---|---|---|---|---|---|
| Article | First Submission Date | Journal | IF | O | Journal | IF | O | Journal | IF | O | Journal | IF | O | Journal | IF | O | Journal | IF | O | Published Date | Att. | Days to publish | Total Times Cited | Citations per year | Article IF (ISI) |
| **Published articles** | | | | | | | | | | | | | | | | | | | | | | | | | |
| Dunn, Moore, & Nosek (2005) | 1/27/2005 | ASAP | U | A | | | | | | | | | | | | | | | | 5/2/2005 | 1 | 95 | 19 | 2.7 | |
| Nosek, Sriram, & Umansky (2012) | 1/18/2012 | PLoS ONE | 4.4 | A | | | | | | | | | | | | | | | | 5/5/2012 | 1 | 108 | 0 | 0 | |
| Vilathong T., Lindner, & Nosek (2010) | 2/24/2010 | JSSR | 1.3 | A | | | | | | | | | | | | | | | | 9/1/2010 | 1 | 189 | 1 | 0.5 | |
| Schwartz, Vartanian, Nosek, & Brownell (2006) | 7/25/2005 | Obesity | 3.5 | A | | | | | | | | | | | | | | | | 3/1/2006 | 1 | 219 | 126 | 21.0 | 7.5 |
| Greenwald et al. (2009) | 4/3/2009 | ASAP | U | A | | | | | | | | | | | | | | | | 11/24/2009 | 1 | 235 | 31 | 10.3 | 6.0 |
| Schmidt & Nosek (2010) | 7/5/2009 | JESP | 2.2 | A | | | | | | | | | | | | | | | | 3/1/2010 | 1 | 239 | 13 | 6.5 | |
| Greenwald, Nosek, & Banaji (2003) | 10/31/2002 | JPSP | 5.2 | A | | | | | | | | | | | | | | | | 8/1/2003 | 1 | 274 | 1375 | 152.8 | 27.5 |
| Nosek & Banaji (2001) | 1/15/2001 | Social Cognition | 1.8 | A | | | | | | | | | | | | | | | | 11/1/2001 | 1 | 290 | 432 | 39.3 | 7.5 |
| Nosek et al. (2009) | 9/10/2008 | Science | 31.4 | R | Nature | 36.1 | R | PNAS | 9.8 | A | | | | | | | | | | 6/30/2009 | 3 | 293 | 51 | 17.0 | 7.0 |
| Ranganath & Nosek (2008) | 3/25/2007 | PS | 4.7 | A | | | | | | | | | | | | | | | | 3/1/2008 | 1 | 342 | 27 | 6.8 | 5.0 |
| Hofmann, Gschwendner, Nosek, & Schmitt (2005) | 12/13/2004 | ERSP | 2.0 | A | | | | | | | | | | | | | | | | 12/1/2005 | 1 | 353 | 94 | 13.4 | |
| Nosek et al. (2010) | 9/14/2009 | Psych. Methods | 3.2 | R | JPSP | 5.2 | R | PSPB | 2.5 | A | | | | | | | | | | 10/1/2010 | 3 | 382 | 2 | 1.0 | |
| Graham, Haidt, & Nosek (2009) | 3/13/2008 | JPSP | 5.2 | A | | | | | | | | | | | | | | | | 5/1/2009 | 1 | 414 | 181 | 60.3 | 26.5 |
| Peris, Teachman, & Nosek (2008) | 8/7/2007 | JNMD | 1.8 | A | | | | | | | | | | | | | | | | 10/1/2008 | 1 | 421 | 22 | 5.5 | 5.5 |
| Lindner & Nosek (2009) | 11/5/2007 | Political Psych. | 1.6 | A | | | | | | | | | | | | | | | | 1/1/2009 | 1 | 423 | 5 | 1.7 | 1.0 |
| Haeffel et al. (2007) | 4/3/2006 | BRAT | 3.0 | A | | | | | | | | | | | | | | | | 6/1/2007 | 1 | 424 | 42 | 8.4 | 3.0 |
| Houben, Nosek, & Wiers (2010) | 10/10/2008 | DAD | 3.4 | A | | | | | | | | | | | | | | | | 1/15/2010 | 1 | 462 | 7 | 3.5 | |
| Kugler, Cooper, & Nosek (2010) | 3/2/2009 | PSPR | 6.1 | R | JPSP | 5.2 | R | SJR | 1.0 | A | | | | | | | | | | 9/1/2010 | 3 | 548 | 7 | 3.5 | |
| Joy-Gaba & Nosek (2010) | 2/15/2009 | Social Psych. | 1.0 | A | | | | | | | | | | | | | | | | 9/1/2010 | 1 | 563 | 13 | 6.5 | |
| Nosek, Banaji, & Greenwald (2002) | 6/16/2000 | GD | 0.9 | A | | | | | | | | | | | | | | | | 1/1/2002 | 1 | 564 | 495 | 49.5 | 18.0 |
| Nosek, Greenwald, & Banaji (2005) | 7/8/2003 | JPSP | 5.2 | R | PSPB | 2.5 | A | | | | | | | | | | | | | 2/1/2005 | 2 | 574 | 359 | 51.3 | 25.5 |
| Bar-Anan, De Houwer, & Nosek (2010) | 3/2/2009 | JEP:LMC | 2.8 | R | QJEP | 2.2 | A | | | | | | | | | | | | | 12/1/2010 | 2 | 639 | 10 | 5.0 | |
| Nosek, Banaji, & Greenwald (2002) | 9/26/2000 | JPSP | 5.2 | A | | | | | | | | | | | | | | | | 7/1/2002 | 1 | 643 | 337 | 33.7 | 13.0 |
| Nosek (2005) | 1/17/2004 | JPSP | 5.2 | R | JEP:G | 5.0 | A | | | | | | | | | | | | | 11/1/2005 | 2 | 654 | 302 | 43.1 | 19.5 |
| Graham et al. (2011) | 4/15/2009 | JPSP | 5.2 | A | | | | | | | | | | | | | | | | 2/1/2011 | 1 | 657 | 21 | 21.0 | |
| Ranganath, Smith, & Nosek (2008) | 5/9/2006 | JESP | 2.2 | A | | | | | | | | | | | | | | | | 3/1/2008 | 1 | 662 | 50 | 12.5 | 4.0 |
| Bar-Anan, Nosek, & Vianello (2009) | 11/14/2007 | JESP | 2.2 | R | EP | 2.1 | A | | | | | | | | | | | | | 10/1/2009 | 2 | 687 | 29 | 9.7 | 6.0 |
| Nosek & Smyth (2011) | 10/1/2009 | JEP | 3.6 | R | AERJ | 2.5 | A | | | | | | | | | | | | | 9/12/2011 | 2 | 711 | 3 | 3.0 | |
| Greenwald et al. (2002) | 1/19/2000 | Psych. Review | 7.8 | A | | | | | | | | | | | | | | | | 1/1/2002 | 1 | 713 | 802 | 80.2 | 24.0 |
| Ratliff & Nosek (2011) | 12/1/2009 | PSPB | 2.5 | A | | | | | | | | | | | | | | | | 12/1/2011 | 1 | 730 | 0 | 0.0 | |
| Nosek & Smyth (2007) | 7/21/2004 | PS | 4.7 | R | PSPB | 2.5 | R | JESP | 2.2 | R | EP | 2.1 | A | | | | | | | 1/1/2007 | 4 | 894 | 104 | 20.8 | 9.5 |
| Mitchell, Nosek, & Banaji (2003) | 11/7/2000 | JPSP | 5.2 | R | JEP:G | 5.0 | A | | | | | | | | | | | | | 9/1/2003 | 2 | 1028 | 210 | 23.3 | 10.5 |
| Ratliff & Nosek (2010) | 11/3/2007 | EP | 2.1 | R | JESP | 2.2 | A | | | | | | | | | | | | | 9/1/2010 | 2 | 1033 | 7 | 3.5 | |
| Sabin, Nosek, Greenwald, & Rivara (2009) | 8/15/2006 | Medical Care | 3.2 | R | SSM | U | R | JHCPU | 1.0 | A | | | | | | | | | | 8/20/2009 | 3 | 1101 | 27 | 9.0 | 6.0 |
| Sriram, Greenwald, & Nosek (2010) | 8/10/2006 | Psych. Methods | 3.2 | R | Psychometrika | 1.8 | R | SM | U | A | | | | | | | | | | 5/1/2010 | 3 | 1360 | 5 | 2.5 | |
| Smith & Nosek (2011) | 8/9/2007 | BJSP | 2.1 | R | Social Cognition | 1.8 | R | Social Psych. | 1.0 | A | | | | | | | | | | 12/1/2011 | 3 | 1575 | 13 | 13.0 | |
| Nosek & Hansen (2008) | 1/4/2004 | JPSP | 5.2 | R | JPSP | 5.2 | R | JEP:G | 5.0 | R | CE | 2.1 | A | | | | | | | 6/1/2008 | 4 | 1610 | 59 | 14.8 | 8.0 |
| Smith, Ratliff, & Nosek (2012) | 7/13/2007 | JESP | 2.2 | R | PSPB | 2.5 | R | Social Cognition | 1.8 | A | | | | | | | | | | 4/1/2012 | 3 | 1724 | 0 | 0.0 | |
| Uhlmann & Nosek (2012) | 3/31/2005 | JESP | 2.2 | R | CE | 2.1 | R | Social Psych. | 1.0 | A | | | | | | | | | | 4/1/2012 | 3 | 2558 | 0 | 0.0 | |
| **Accepted but not yet published** | | | | | | | | | | | | | | | | | | | | | | | | | |
| Hunt, Gonsalkorale, & Nosek (2012) | 2/7/2011 | IJMH | U | A | | | | | | | | | | | | | | | | . | 1 | 451 | | | |
| Hawkins & Nosek (2012) | 10/20/2010 | JPSP | 5.2 | R | PNAS | 9.8 | R | Science | 31.4 | R | PS | 4.7 | R | PSPB | 2.5 | A | | | | . | 5 | 561 | | | |
| Menatti, Smyth, Teachman, & Nosek (2012) | 10/6/2010 | SPPS | U | R | JSCP | 1.3 | R | JNMD | 1.8 | R | BASP | 0.7 | R | GPIR | 1.4 | R | SRA | U | A | . | 6 | 575 | | | |
| Ratliff, Swinkels, Klerx, & Nosek (2012) | 9/6/2010 | JCR | 2.4 | R | PM | 1.4 | A | | | | | | | | | | | | | . | 2 | 605 | | | |
| Bar-Anan & Nosek (2012) | 8/10/2009 | JPSP | 5.2 | R | JESP | 2.2 | R | PSPB | 2.5 | R | Social Cognition | 1.8 | R | CE | 2.1 | R | PSPB | 2.5 | A | . | 6 | 997 | | | |
| | | | | | | | | | | | | | | | | | | | | | | | | | |
| **Not yet accepted** | | | | | | | | | | | | | | | | | | | | | | | | | |
| Sabin, Marini, & Nosek | 1/22/2012 | Obesity | 3.5 | R | | | | | | | | | | | | | | | | . | 1 | 102 | | | |
| Marini, et al. | 1/12/2012 | Science | 31.4 | R | Nature | 36.1 | R | PNAS | 9.8 | R | | | | | | | | | | . | 3 | 112 | | | |
| Nosek, Bar-Anan, Sriram, & Greenwald | 1/1/2012 | JESP | 2.2 | R | | | | | | | | | | | | | | | | . | 1 | 123 | | | |
| Friese, Smith, Plischke, Bluemke, & Nosek | 12/12/2011 | JApP | 4.0 | R | JEP:G | 5.0 | R | PLoS ONE | 4.4 | U | | | | | | | | | | . | 3 | 143 | | | |
| Smith & Nosek | 8/12/2011 | JPSP | 5.2 | R | | | | | | | | | | | | | | | | . | 1 | 265 | | | |
| Smith, De Houwer, & Nosek | 7/14/2011 | JESP | 2.2 | R | PSPB | 2.5 | U | | | | | | | | | | | | | . | 2 | 294 | | | |
| Lindner, Graser, & Nosek | 2/14/2011 | PA | 3.1 | R | JASP | 0.7 | U | | | | | | | | | | | | | . | 2 | 444 | | | |
| Graham, Sherman, Iyer, Hawkins, Haidt, & Nosek | 12/15/2010 | Science | 31.4 | R | PNAS | 9.8 | R | Cognition | 3.7 | R | JEP:G | 5.0 | R | | | | | | | . | 4 | 505 | | | |
| Riskind, Nosek, & Patterson | 12/10/2010 | PS | 4.7 | R | JFP | 1.9 | R | HR | 4.4 | R | JSPR | 1.0 | R | Parenting | U | R | | | | . | 5 | 510 | | | |
| Hawkins & Nosek | 8/16/2010 | PSPB | 2.5 | R | BJSP | 2.1 | R | PloS ONE | 4.4 | U | | | | | | | | | | . | 3 | 626 | | | |
| Smyth, Greenwald, & Nosek | 6/29/2009 | SPPS | U | R | JESP | 2.2 | R | | | | | | | | | | | | | . | 2 | 1039 | | | |
| Graham, Nosek, & Haidt | 12/18/2008 | Political Psych. | 1.6 | R | PSPB | 2.5 | R | JESP | 2.2 | U | | | | | | | | | | . | 3 | 1232 | | | |
| Bar-Anan & Nosek | 6/26/2007 | JPSP | 5.2 | R | JESP | 2.2 | R | | | | | | | | | | | | | . | 2 | 1773 | | | |
| Devos, Nosek, & Banaji | 3/30/2007 | GPIR | 1.4 | R | | | | | | | | | | | | | | | | . | 1 | 1861 | | | |
| Sriram, Nosek, & Greenwald | 11/28/2006 | Psychometrika | 1.8 | R | Psych. Review | 7.8 | R | | | | | | | | | | | | | . | 2 | 1983 | | | |
| Friedman et al. | 12/1/2001 | JAbP | 5.2 | R | | | | | | | | | | | | | | | | . | 1 | 3806 | | | |
| Nosek | 10/24/2001 | PS | 4.7 | R | | | | | | | | | | | | | | | | . | 1 | 3844 | | | |
| Nosek & Ahn | 7/15/1999 | PS | 4.7 | R | | | | | | | | | | | | | | | | . | 1 | 4676 | | | |
| | | | | | | | | | | | | | | | | | | | | | | | | | |
| **Summary** | | | | | | | | | | | | | | | | | | | | | | | | | |
| Published articles | | Median IFs | 3.2 | | | 2.5 | | | 1.8 | | | 2.1 | | | . | | | . | | Means | 1.7 | 677 | 135 | 19.4 | 11.5 |
| Accepted but not yet published | | | 3.2 | | | 2.5 | | | 2.0 | | | 2.1 | | | 2.1 | | | 2.5 | | | 4.0 | 638 | . | . | . |
| Not yet accepted | | | 4.0 | | | 2.5 | | | 4.4 | | | 3.0 | | | . | | | . | | | 2.1 | 1297 | . | . | . |
| Total | | | 3.4 | | | 2.5 | | | 2.5 | | | 2.1 | | | 2.1 | | | 2.5 | | | 2.0 | 854 | | | |

Table 2. Correlations among indicators of the publishing process and article impact for the case study

| **Correlations** | **Total Times Cited** | **Citations per year** | **Article IF** | **First journal IF** | **Publishing journal IF** |
|---|---|---|---|---|---|
| Sample size | 33 | 33 | 21 | 30 | 30 |
| Date of publication | 0.71* | 0.61* | 0.50* | -0.02 | 0.25 |
| Number of journals attempted | -0.22 | -0.20 | -0.12 | 0.34 | -0.06 |
| Days between first submission to publication | -0.10 | -0.10 | -0.05 | -0.07 | -0.12 |
| Impact factor (IF) of first journal | 0.05 | 0.11 | 0.04 | | 0.75* |
| Impact factor (IF) of publishing journal | 0.35* | 0.42* | 0.35 | 0.75* | |

Note: * p < .05; Includes articles in print for at least one year (two years for article IF). Total times cited and citations per year data from Google Scholar; Article IF, First journal IF and Publishing journal IF data from ISI (when available)